\documentclass[aps,pra,reprint,groupedaddress,showpacs]{revtex4-1}

\usepackage{graphics}
\usepackage{hyperref}
\newcommand{\bi}{\begin{itemize}}
\newcommand{\ei}{\end{itemize}}
\newcommand{\be}{\begin{equation}}
\newcommand{\ee}{\end{equation}}
\newcommand{\bea}{\begin{eqnarray}}
\newcommand{\eea}{\end{eqnarray}}
\newcommand{\beas}{\begin{eqnarray*}}
\newcommand{\eeas}{\end{eqnarray*}}

\newcommand{\bfr}{{\bf r}}

\newcommand{\bfu}{{\bf u}}

\newcommand{\Exc}{ E_{xc} }

\newcommand{\gcbar}{\bar{g}_c}
\newcommand{\ncbar}{\bar{n}_c}

\newcommand{\sysnxc}{ \langle n_{xc}( u ) \rangle  }
\newcommand{\sysnx}{ \langle n_{x}( u ) \rangle  }
\newcommand{\sysnc}{  \langle n_{c}( u ) \rangle  }

\newcommand{\avrs}{ \langle r_s \rangle }
\newcommand{\avrssigma}{ \langle r_s^\sigma \rangle }
\newcommand{\avzeta}{ \langle \zeta \rangle }
\newcommand{\avssq}{ \langle s^2 \rangle }
\newcommand{\avtsq}{ \langle t^2 \rangle }
\newcommand{\avt}{ t_{rms} }
\newcommand{\avs}{ s_{rms} }
\newcommand{\tilders}{ \tilde{r}_s  }
\newcommand{\tildezeta}{ \tilde{ \zeta } }
\newcommand{\tildessq}{ \tilde{s}^2 }
\newcommand{\tildetsq}{ \tilde{t}^2  }

\newcommand{\nup}{ n_{\uparrow}(\bfr) }
\newcommand{\ndn}{ n_{\downarrow}(\bfr) }

\newcommand{\dnup}{ \nabla n_{\uparrow}(\bfr) }
\newcommand{\dndn}{ \nabla n_{\downarrow}(\bfr) }

\begin{document}

\preprint{Cancio, preprint 2012}

\title{The scaling properties of exchange and correlation holes of the valence shell of second row atoms}


\author{Antonio C. Cancio}
\affiliation{Ball State University, Muncie, IN 47306}
\email[]{accancio@bsu.edu}
\author{C. Y. Fong}
\affiliation{Department of Physics, University of California, Davis, CA 95616}


\date{\today}

\begin{abstract}
We study the exchange and correlation hole of the 
valence shell of second row atoms using variational Monte Carlo techniques,
especially correlated estimates, and norm-conserving pseudopotentials.
The well-known scaling of the valence shell provides a tool to probe
the behavior of exchange and correlation as a functional of the density and 
thus test models of density functional theory.
The exchange hole shows an interesting competition between two scaling 
forms -- one caused by self-interaction and another that is approximately 
invariant under particle number, related to the known invariance of 
exchange under uniform scaling to high density and constant particle number.
The correlation hole shows a scaling trend that is marked by the finite
size of the atom relative to the radius of the hole.
Both trends are well captured in the main by the Perdew-Burke-Ernzerhof 
generalized-gradient approximation model for the exchange-correlation hole 
and energy.
\end{abstract}

\pacs{31.15.ae, 71.15.Mb, 02.70.Ss}

\maketitle

\makeatletter
\global\@specialpagefalse
\def\@oddhead{REV\TeX{} 4.0\hfill Cancio, Preprint, 2012}
\let\@evenhead\@oddhead
\makeatother


\section{Introduction}

Density functional theory~\cite{HK,JG}, the most widely used computational tool for electronic structure
calculations, is founded upon the knowledge of the existence of a universal functional mapping the 
ground-state density to the ground-state energy, with, however,
a fundamental lack of 
knowledge how to construct this functional systematically.  
The key problem is describing the effects of electron-electron 
interactions due to Fermi statistics and Coulomb repulsion -- 
 the exchange and correlation energies -- in an essentially single-particle 
description of nature. 
Considerable progress in constructing approximate functionals has been made,
using a number of varied strategies to develop a
``Jacobs ladder" hierarchy of models of increasing complexity. 
Two such strategies which have proved fruitful are the discovery and implementation
of scaling laws that describe limiting behavior of the universal density functional and the analysis of auxiliary expectations, particularly the exchange-correlation  hole, to provide insight into 
the role of interelectron correlations in determining this functional.


In DFT, scaling laws provide a controlled way
of varying density, approaching the daunting task of understanding
the energy as a functional of the density  
by first tackling the more approachable one of understanding its behavior as
a function of a scaling variable.  
Particularly useful is the limit of uniform scaling to high 
density~\cite{LevyPerdew,Levy}, a process intimately related to the adiabatic 
connection approach to DFT~\cite{Exc2nxc}, and asymptotically approached by the 
isoelectronic series of atomic ions as nuclear charge $Z$ tends to infinity.  
The properties of exchange and correlation under this tranformation 
constrain the possible dependence of the energy on local-density based 
variables, greatly simplifying the task of functional construction.
Another fruitful 
scaling process is that of neutral atoms in the $Z\! \rightarrow \!\infty$ limit, 
in which the charge density tends to a well known Thomas-Fermi 
limit~\cite{Scott,MarchTF}.  The gradient correction of the latter has been 
used to diagnose a major limitation in the widely used generalized gradient 
approximation (GGA) of DFT, namely the limited ability to tune the GGA to 
predict accurately both molecular and solid-state properties~\cite{PCSB-PRL97}, and to 
motivate effective recent remedies for this issue~\cite{PBEsol,revTPSS}.  

Another important tool in the development of DFT has been 
the exchange-correlation (XC) hole.
The XC hole essentially is the measure of the change in electron number 
density throughout a system given one electron observed to be at a given position.
The energy of interaction of an electron with its own exchange-correlation
hole yields the exchange-correlation energy, the key theoretical input
into DFT. 
The unexpected degree of success of the original local density approximation (LDA) stems from the 
universal properties of the system-averaged XC hole (a hole averaged
over angle and position in the system) obeyed by the homogeneous electron gas 
hole~\cite{JG} from which the LDA is derived.
Input from the behavior of the XC hole has been important in the development
of effective nonempirical GGA's~\cite{PW91,PBW_GGA,PBE} and the 
hybrid DFT-Hartree Fock approach~\cite{BeckeHybrid}.
More recent DFT models   
have not been constructed from XC holes, 
but constructing model XC holes consistent with a given functional remains an 
important analysis tool~\cite{TPSShole}.

The system-averaged exchange-correlation hole
has a natural connection to the 
intracule~\cite{Coulson} or density of electron-pairs as a function of their
interelectron distance.  The intracule has been the subject of extensive study in quantum chemistry,
as a source of insight into the electron correlation problem.  
Several classes of techniques have been used to study atoms and simple molecules, including 
Hartree Fock~\cite{KogaMatsu,MKRD,Koga1},
configuration interaction methods~\cite{BanyardMobbs,FDM,FDVU,WTS1,WTS2,CiosLiu2,*CiosLiu3},
and quantum Monte Carlo (QMC)~\cite{CancioFong,Galvez:034302,*Galvez:044319,Toulouse}.
Among the many applications include
scaling across isoelectronic series~\cite{Galvez:034302,Galvez:044319},
scaling across the periodic table~\cite{MKRD,KogaMatsu,Koga1}, decomposition into
approximate exchange and correlation components~\cite{FDM,WTS1,WTS2} and
shell analysis~\cite{BanyardMobbs,MKRD}.
Pair densities are of additional interest because they can be measured experimentally~\cite{Thakkar,WTS2} and 
are the basis of some density functional theory approaches~\cite{GoriSavin,Nagy}.

The decomposition of the intracule into Fermi, or Hartree-Fock hole, and 
Coulomb hole, incorporating effects from correlations, 
is a close but imperfect equivalent to exchange and correlation in DFT.  
The former uses the Hartree-Fock ground-state 
density as the reference point for defining both Fermi and Coulomb correlations 
rather than the exact ground-state density as required in DFT.  This 
difference produces a failure of the virial theorem at the level of correlation 
that can be a ten percent effect in the case of an atom or a major issue
in the case of a molecule near dissociation~\cite{HoleReview}.  
Secondly, standard Coulomb holes account only for the correlation potential 
energy.  To obtain the correlation contribution to the kinetic energy -- the 
difference between the true and noninteracting-system kinetic energies -- an 
adiabatic integration of the correlation hole with respect to coupling-constant 
must be performed~\cite{Exc2nxc}.  Calculations of ``true" exchange-correlation 
holes, using orbitals that reproduce at least the density of one's correlated 
wavefunction, have been done mainly in the QMC 
approach~\cite{FWLnxc,Hood1,Hood2,NekoveeAll,Cancio,Puzder,Hsing}.

In this paper, we calculate and analyze the exchange and correlation holes of the valence shell of 
second row atoms in a pseudopotential model using the variational quantum Monte Carlo (VMC) technique.  
To our knowledge, this scaling phenomenon
has not been studied in the context of density functional theory -- 
although it was an important ingredient in the earliest attempts at 
a pseudopotential description of atomic structure~\cite{Slater}. 
Uniform scaling of the radial valence-charge density 
across a row of the periodic table provides a convenient way 
to test the idea of ``semilocality" of the system-averaged hole by
varying the ratio of correlation length to atomic radius
over a series of otherwise similar systems.
The second row of the periodic table is easy to handle with pseudopotential methods and 
thus easy to isolate strictly valence properties.  The valence density 
that results is very close to scale-invariant.
The systems studied are building blocks of semiconductor materials, and
probe a density regime, lower than that of the 1st row valence shell, 
typical of many solids for which pseudopotential simulations are generally used. 

The variational Monte Carlo approach~\cite{Foulkes} makes feasible the use of explicitly-correlated 
trial wavefunctions, including the electron-electron cusp condition that affects the 
pair-density at zero electron-pair separations.  
To obtain separate system-averaged exchange and correlation holes, 
we construct single-particle orbitals that reproduce the VMC single-particle 
density.  Exchange holes are then calculated numerically
from these orbitals.  To reduce the errors from fluctuations in the 
pair density due to random sampling, correlated estimates techniques are used.
These prove important for the measurement of the correlation 
hole which is a small fraction of the total pair density and thus more 
affected by noise.  The resulting 
exchange and correlation holes are analyzed with respect to the scaling of the 
valence shell density across the row and with respect to various
density functional models.  

The paper is organized as follows.  Section~\ref{section:theory} discusses the theoretical 
underpinnings of the paper -- exchange and correlation holes, 
the GGA approximation,
and the scaling of the valence shell. 
Section~\ref{section:methods} describes the computation techniques used to generate holes and
other expectations, Section~\ref{section:results} presents our results and Section~\ref{section:conclusions} our conclusions.  
All results are expressed in hartree atomic units.


\section{Theory}
\label{section:theory}
\subsection{Expectations of interest}


The exchange-correlation (XC) hole, 
\mbox{$n_{xc}(\bfr, \bfr + \bfu)$},
is defined as the reduction in the ground-state electron density 
from its mean value at some point $\bfr \!+\! \bfu$ given the observation
of an electron at $\bfr$.  
It is obtained from a pair density fluctuation relationship:
\be
     n(\bfr) n_{xc}(\bfr, \bfr + \bfu) = 
         n^{(2)}(\bfr, \bfr + \bfu) -  n(\bfr + \bfu)n(\bfr)
     \label{eq:nxc}
\ee
where $n(\bfr)$
is the single-particle density, and
\begin{equation}
     n^{(2)}({\bf r},{\bf r'}) =  \left\langle\sum_i^N\sum_{j\neq i}^N 
             \delta({\bf r} - {\bf r_i}) \delta({\bf r'} - {\bf r_j})\right\rangle.  
     \label{eq:pairdens}
\end{equation}
is the pair density, measuring the expectation of simultaneously finding 
electrons at $\bfr$ and $\bfr'$.  
Eq.~(\ref{eq:nxc}) relates the XC hole to the difference
between the actual pair density and that of the uncorrelated system with
the same single-particle density.

The utility of the XC hole in density functional theory 
lies in its relation to the 
exchange-correlation energy $E_{xc}$ through an adiabatic connection~\cite{Exc2nxc}:
        \be
           E_{xc} = 
             \frac{1}{2}\displaystyle\int_0^1 d\lambda 
           \displaystyle\int d^3r \displaystyle\int d^3u 
           \; \frac{1}{u} n^{\lambda}_{xc}(\bfr, \bfr + \bfu).
        \ee
$\Exc$ takes into account both the gain in potential energy from 
creating the exchange-correlation hole about an electron, and, 
by means of the integration over the coupling constant $\lambda$, 
the kinetic energy cost of creating the hole as well.
The $\lambda$ integral is over a family of systems characterized by the same 
ground-state density but varying coupling constant $\lambda e^2$. 
A $\lambda$-dependent XC hole $n^\lambda_{xc}$ is defined as an 
expectation of the corresponding ground-state wavefunction.
The limits of the integral range from a noninteracting system ($\lambda\!=\!0$),
described by the Kohn-Sham equations of DFT~\cite{KohnSham}, 
to the fully interacting, physical system ($\lambda\!=\!1$).

The $\lambda\!=\!1$ limit which describes the fully 
interacting Hamiltonian is the 
focus of this paper.  By ignoring the integration over coupling constant,
we lose the ability to calculate the correlation kinetic energy 
and thus lose some of the information contained in the full XC energy.
However, we keep the ability to assess DFT models for this quantity
since one can use a scaling relation to 
convert the hole integrated over $\lambda$
into that evaluated at any specific value of $\lambda$. 
This thereby eliminates a tedious chore for the quantum
Monte Carlo method 
and allows one to explore multiple systems more 
readily~\cite{AdiabaticFootnote}.

As the Coulomb interaction depends only on the interparticle
distance $u$, and not on the location or angular orientation of the hole,
it is convenient to define a system- and angle-averaged hole, 
\be
\sysnxc  =  \frac{1}{4\pi} \int \!d^3r\! \int\! d\Omega_u\; n(\bfr) 
       n_{xc}(\bfr,\bfr + \bfu)\;
\label{eq:sysavnxc}
\ee
where $d\Omega_u$ is the solid angle of the pair displacement $\bfu$.
This expression
may be normalized by the number of particles $N$ or number of particle pairs
$N^2$.
The system-averaged hole contains that part of the XC hole that directly
affects the determination of $E_{xc}$, which 
simplifies to
        \be
           E_{xc} = 
               \displaystyle\int_0^1 \!d\lambda E_{xc}^{\lambda} =       
               \displaystyle\int_0^1 \!d\lambda 
               \displaystyle\int \!du \;2\pi u \langle n^{\lambda}_{xc}(u)\rangle
           \label{eq:exc}
        \ee
because of the isotropy of the Coulomb interaction.
The system-averaged hole is closely related to 
the intracule, defined as:
\be
\langle n^{(2)}(u)\rangle = 
        \left\langle \sum_{i\neq j} \delta(u - r_{ij}) \right\rangle. 
\label{eq:intracule}
\ee
The system-averaged hole is the difference between the
intracule of a quantum system and that of the uncorrelated
system with the same single-particle density.
%

The XC hole is usefully decomposed in
several ways to isolate various sources of electron correlation.
The exchange (X) hole $n_x$ is defined as the exchange-correlation hole
at zero coupling or $n_{xc}^{\lambda\!=\!0}$.
This is the hole associated with the Slater determinant
wavefunction that reproduces the ground-state density of the fully
interacting system and characterizes the inter-electron correlations
due to the Pauli principle.
The correlation (C) hole is the difference
between the exchange and exchange-correlation holes 
and measures the additional many-body
correlations induced by the Coulomb interaction:
\be
      \sysnc = \sysnxc - \sysnx.
      \label{eq:nc}
\ee

In addition, it is useful to consider 
the spin-decomposition of the XC hole, particularly in spin-polarized
systems.  One may define parallel and anti-parallel-spin holes by
restricting the sums in Eq.~(\ref{eq:pairdens}) 
to particles with the same or opposite spins respectively.
The parallel spin hole is dominated by the exchange contribution,
since the exchange hole already creates distance between electrons
of the same-spin, so that further effects due to correlation are small.
On the other hand, the antiparallel spin channel, where
the exchange hole is zero, contributes the bulk of the correlation hole.

As it describes the noninteracting system, the spin-decomposed exchange-only hole
can be written exclusively 
in terms of the Kohn-Sham single-particle orbitals that define this limit.
The system-averaged hole in this case is obtained exactly as
\be
    \langle n^{\sigma\sigma}_x(u)\rangle = 
          -\int d^3r \int \frac{\Omega_u}{4\pi}
          \left|\displaystyle\sum_{i=1}^{N_\sigma} 
                 \psi_{i\sigma}(\bfr) \psi^*_{i\sigma}(\bfr+\bfu) \right|^2
     \label{eq:sysavnxsigma}
\ee
where 
\be
      \langle n_x(u) \rangle = 
            \sum_\sigma \langle n^{\sigma\sigma}_x(u) \rangle,
      \label{eq:nxone}
\ee
and $\psi_{i\sigma}$ are Kohn-Sham orbitals for each spin.




Finally worth noting is that the XC hole obeys the
sum rule:
    \be
	\frac{1}{N} \int 4\pi u^2 du \sysnxc = -1.
    \ee
The overall effect of the hole is to
remove exactly one particle from the measurement of the 
density about any given electron, essentially removing self-interaction.
Given their respective definitions, 
the exchange hole must satisfy the same sum rule as exchange-correlation, 
while the correlation hole, being merely a redistribution of the $N-1$
 other electrons around the one in consideration, integrates to zero.

\subsection{Exchange-correlation Hole in Semilocal Density Functional
Theory}

In a ``semilocal" density functional theory, the exchange-correlation hole
at some point $\bfr$ in space
is constructed in terms of various observables defined 
at that point: local spin densities 
$\nup\!$ and $\ndn\!$ in the local spin-density (LSD) variant of the LDA~\cite{LSD,JG}, and
adding density gradients $\dnup\!$ and $\dndn\!$ in generalized gradient
approximations (GGA's)~\cite{PBE,BeckeGGA,LYP,MattssonGGA,SOGGA}.  
Kinetic energy densities~\cite{BeckemGGA,TPSS,SchmiderBecke,M06L}, 
and higher-order derivatives of the density such as 
the Laplacian~\cite{LYP, PerdewConstantin, CancioLapl} are included
in the meta-GGA class of theories.

The  LSD exchange-correlation 
hole at a point $\bfr$ 
is obtained from the pair correlation
function $g_{xc}$ of the spin-polarized homogeneous electron gas (HEG):
\be
             n^{LSD}_{xc}(\bfr, \bfr+\bfu) = 
	     n(\bfr)\{g^{HEG}_{xc}\;[ u,\, r_s(\bfr),\, \zeta(\bfr) ]-1\}.
             \label{eq:nxclsd}
\ee
The pair correlation function is parametrized in terms of the
Wigner-Seitz radius $r_s$, measuring the average distance between electrons, and the 
spin polarization $\zeta$, and both are evaluated from the local value of the 
spin densities:
\bea
       r_s(\bfr) &=& \left( \frac{3} {4\pi n(\bfr)} \right)^{1/3}  \\
       \zeta(\bfr) &=& \frac{\nup\!-\!\ndn} {n(\bfr)}. 
       \label{eq:zeta}
\eea
The system-averaged hole and total XC energy may then be obtained numerically
by applying Eq.~(\ref{eq:sysavnxc}) and Eq.~(\ref{eq:exc}) respectively.


Among the many 
variants of the GGA in current use, one of particular
interest here is that~\cite{PBE} of Perdew, Burke and Ernzerhof (PBE), 
for which models 
of $\sysnx$~\cite{PBW_GGA,EP_Xhole} and $\sysnc$~\cite{PBW_GGA} have been constructed, building upon an accurate HEG hole~\cite{PW}.
Holes within the PBE model are designed, under integration,
to reproduce PBE exchange and correlation energies at any 
value of the density and its gradient within an error of about 5\%.  

For the exchange hole and energy, the PBE and related models introduce 
into Eq.~(\ref{eq:nxclsd}) a unitless, scale-invariant parameter $s$, defined as
\be
             s(\bfr) =  \frac{1} {2 k_F(\bfr)} 
                           \left| \frac{\nabla n(\bfr)} {n(\bfr)} \right|,
\ee
with $k_F = (9\pi/4)^{1/3}/r_s$ the fermi wavevector.  
A value of $s$ greater than one at a given point indicates the breakdown
of the basic assumption of the LSD that the density varies insignificantly
on the length scale $\sim 1/k_F$ of the XC hole.
For correlation, the PBE employs a second inhomogeneity parameter~\cite{PBE},
\be
             t(\bfr) =  \left[ k_F(\bfr)/\phi(\zeta) k_s(\bfr) \right] s(\bfr), 
\ee
with 
the local Thomas-Fermi screening vector, $k_s\! =\! (4 k_{F}(\bfr)/\pi)^{1/2}$, 
setting the hole length scale
and $\phi(\zeta)$, an additional scaling factor for spin-polarized systems.

As the quantity of interest in this paper is the system averaged 
hole rather than the local hole, it is helpful to consider
system-averages of semilocal DFT parameters.
A simple definition of the system-averaged Wigner radius $\avrs$ for an 
atom is~\cite{BPE-ontop}
\be
\avrs = \displaystyle\frac{ \int dr\: r^2 n(r)^2 r_s(r) }
			  { \int dr\: r^2 n(r)^2 },
\label{eq:avrs}
\ee
assuming the pair density for zero interparticle separation~\cite{FDVU} is a reasonable measure of 
the importance of the hole at $\bfr$ to energy expectations, and ignoring
the anisotropy of the density.
System-averaged spin polarization $\avzeta$ and
inhomogeneity factors $\avssq$ and $\avtsq$ are similarly defined.

Finally, it is important to note that 
we are calculatiing the system-averaged correlation hole at full coupling 
constant ($\lambda \!=\! 1$), and with it, the correlation potential energy.  
DFT models for the correlation hole 
model the adiabatically integrated hole and thus the total correlation energy.  
It is possible however to construct the DFT correlation hole and energy density 
at a given value of coupling constant $\lambda$ from the adiabatically 
integrated version. 
This can be derived from the scaling properties of the hole under
uniform scaling of the system~\cite{PW}.
For the GGA the following expression for the pair correlation function
is the result:
\be
   g_c^{1}(r_s, s, \zeta, k_Fu) = 
         \gcbar( r_s, s, \zeta, k_Fu) - 
         \frac{\partial \gcbar(r_s, s, \zeta, k_Fu)}{\partial \ln{r_s}},
\label{eq:gcscaling}
\ee
with the correlation hole obtained from the expression
\be
   n_c(\bfr,\bfr+\bfu) = n(\bfr) \left[g_c(\bfr,u) - 1\right],
\ee 
definable for both $n_c^{\lambda}$ and $\ncbar$.
The parameters $k_Fu$, $\zeta$, and $s^2$ (but not $t^2$) 
are invariant
under uniform scaling of position coordinates $\bfr_i$ 
(given by $\bfr_i \!\rightarrow\! \bfr_i/\lambda$, and 
$n(\bfr) \!\rightarrow\! n(\bfr/\lambda) / \lambda^3$), and thus held fixed in
the derivative.  
In constrast, the exchange hole is invariant under 
changes in scale or coupling constant, and is constructed in terms of the
scale-independent quantities only.


\subsection{Valence Shell Scaling of Atoms}

A well-known feature of the periodic table is the scaling of the valence-shell
electron density across the 1st or 2nd row atoms 
(the 2s and 2p or the 3s and 3p atoms respectively).  As the number 
of valence electrons $N$ increases, the shell radius $a(N)$ shrinks with
no change in the shape of the distribution:  
   \be
        n(r;N) = \frac{N}{a(N)^3} \bar{n}[r/a(N)].
         \label{eq:scaledens}
   \ee
The scaling behavior strictly occurs for the valence density outside the
core radius.  
But with the use of pseudopotentials to remove
the complicated oscillatory behavior of the valence shell inside the core,
this scaling behavior becomes a global feature of the density -- 
one motive for the development of pseudopotentials historically~\cite{Slater}.
The radius $a$, usefully defined as the radius of the peak in the 
radial distribution function $4\pi r^2 n(r)$, is a function of valence number 
$N$, 
so that, if the distribution
were perfectly scaling, the density should reduce to a function of $N$ 
alone across the row.  
(We consider neutral atoms, with $N=Z_V$, the charge of the ion core.)
As a result 
all other expectations of the pseudopotential system ground-state should be 
reducible to simple functions of $N$, in accordance to the Hohenberg-Kohn 
theorems~\cite{HK}.
Such a picture is limited by the non-scaling behavior of the 
valence density in the ionic core, not a large contribution to the
probability distribution due to the relatively small volume of the core.
It is also affected by differences between the Mg atom and atoms with 3p 
orbitals, and by the open-shell structure of several of the atoms.

%
%

Important insights into this scaling can be gained by the simple heuristic 
shielding model of the atom developed by Slater~\cite{Slater}. 
This model assumes
a self-consistent field felt by an electron in
energy shell $i$ given by a shielded Coulomb potential $(Z-\sigma_i)/r$,
with $\sigma_i$ the shielding charge felt by the electrons of the shell.  
The orbitals for principle quantum number $n$ are nodeless Slater-type
orbitals $r^{n-1} \exp{[(Z-\sigma_i)r/n]}$, imposing
scaling of the density across a row. 
The effective Bohr radius -- the peak of the 
radial probability distribution for the valence shell -- occurs at 
\be
       a = \frac{a_0n^2}{Z-\sigma_V} = \frac{a_0n^2}{A + BN},
       \label{eq:slater}
\ee 
with $\sigma_V$ the valence-shell shielding charge.
Slater's ``back-of-the-envelope"
rules for determining shielding coefficients yield a shielded effective 
charge $Z-\sigma_V$ that is a linear function of $N$, with coefficients for 
the valence shell of the 2nd row atoms of $B\!=\!0.65$ and $A\!=\!1.55$.

Energetic quantities can thus be constructed as functions of $N$ from 
consideration of the scaling form for the density [Eq.~(\ref{eq:scaledens})] and
$a(N)$.  The total energy of the valence shell in the Slater model is 
$(Z - \sigma_V)N/2a(N)$ in hartrees; expressing the shielding
charge in terms of $a(N)$ yielding a scaling dependence of $N/a^2$.  The
kinetic energy in this naive picture scale in the
same way, while the external potential due to the 
ion scales as $N^2/a$ for $Z_V\!=\!N$.  
The Hartree or classic electrostatic energy scales as $N(N-1)/a$ provided
that self-interaction error is removed.  With these scaling assumptions, the
virial theorem relating potential and total energy is satisfiable by
$a(N)$ taking on the Pade function form of Eq.~(\ref{eq:slater})~\cite{SlaterVirial}.
The most interesting energetic
quantity for our purposes is the exchange energy.
The heuristic picture for scaling of valence-shell energies should be 
readily extended to the exchange energy, as it in fact obeys 
an important universal scaling law~\cite{LevyPerdew}: 
\be
E_x[n_{\alpha}] = \alpha E_x[n].
\label{eq:exscaling}
\ee
where the density $n_{\alpha}$ is defined by the uniform scaling
of $n(r)$ at {\it constant particle number}:
\be
n_{\alpha}(r) = \alpha^3 n(\alpha r).
\ee
To touch base with the current situation, we 
note that particle number $N$ is not fixed as one fills
up the valence shell, so that the density scales
with an additional prefactor $N$, with an $N$-dependent 
length scale $a(N)$ playing the role of $1/\alpha$.  
A similar situation is seen in the Thomas
Fermi scaling of all-electron densities of atoms recently revisited in Ref.~\onlinecite{PCSB-PRL97}.  
As with the Hartree energy, and in contrast to the all-electron case,
self interaction is not negligible, varying with particle number and becoming
critical for the smallest systems.
A further complicating factor is that as the shell is filled, the overall 
spin-polarization does not stay constant -- the process of filling the 
shell is not a uniform scaling of the density itself
but at best of the radial component of the density.

To develop scaling model for exchange within DFT, we first construct
scaling expressions for the system-averaged electron-gas
parameters $\avrs$ and $\avzeta$ defined in the previous section.  
The scaling behavior for $\avrs$ is that 
of fitting $N$ particles in a box of volume $a$:  
\be
	\avrs \sim \tilders = R_0 a/N^{1/3},
        \label{eq:scalers}
\ee
with $R_0$ an unknown constant.
This equation with Eq.~(\ref{eq:scaledens}) generates similar scaling relations
for the inhomogeneity parameters $s^2$ and $t^2$, in terms of unknown
constants $S_0$ and $T_0$:
\bea
        \tildessq &=& S_0^2/N^{2/3} \\
        \tildetsq &=& T_0^2/aN^{1/3}. 
        \label{eq:scales2}
\eea
A useful estimate of the system-averaged spin-polarization $\avzeta$ is given by
    \be
        \tildezeta = \frac{ N_{\uparrow} - N_{\downarrow} }{N}.
        \label{eq:scalezeta}
    \ee
Note that $\tildessq$ does not depend on the scaling parameter $a$, but only 
on the number of particles $N$ -- it is invariant under uniform 
scaling of the density but not, of course, under change in $N$.
The polarization $\tildezeta$ is also scale invariant.

A scaling form for the exchange energy can be constructed within the LSD
in terms of the scaled parameters $\tilders$ and $\tildezeta$, 
incorporating the invariance of exchange of the homogeneous electron
gas under uniform scaling of the density, and a similar spin-density scaling 
relationship.  
The scaled version of the exchange energy within the LSD becomes:
    \be
    E_x  \sim \frac{N}{\tilders} \phi_X(\tildezeta),
        \label{eq:scaleex}
    \ee
where
    \be
    \phi_X(\tildezeta) = \frac{1}{2}
            \left[ \left(1 + \tildezeta \right)^{4/3} +
                   \left( 1 - \tildezeta \right)^{4/3}\right].
        \label{eq:phix}
    \ee
This is directly obtainable from the definition of the 
local energy-per-particle
in the LSD~\cite{OliverPerdew}, replacing local definitions for $r_s$ and $\zeta$ with 
system-averaged counterparts.



There is no simple scaling law for correlation to correspond to
that of exchange and the correlation energy is difficult
to model even for the homogeneous electron gas.  
The analog to the exchange scaling of Eq.~(\ref{eq:exscaling}) does exist for 
the correlation potential energy~\cite{Levy}:
\be
U_c^1[n_\alpha] = \alpha U_c^{1/\alpha}[n]
\label{eq:ecscale}
\ee
tying the process in which the density is scaled uniformly by $\alpha^3$ to 
that in which the coupling constant is 
reduced at fixed density~\cite{footnoteUc}.  
This suggests that the Ar correlation hole is equivalent to Mg with 
a electrostatic coupling reduced to allow the binding of the full $p$ shell.  
It does not relate the two systems at full coupling.
In systems where $Z$ is varied at constant particle number, Levy
has shown~\cite{Levy} that $E_c(Z)$ must tend to a constant at large $Z$ 
(that is, uniform scaling to high density).  It is possible to expect some
analogous limiting case for scaling of neutral atoms, 
but this limit is apparently unknown.

\section{System and Calculation Methods:}
\label{section:methods}
\subsection{Hamiltonian}
Our system of interest is described by a  
many-body Hamiltonian for $N$ valence electrons, with a nonlocal
pseudopotential~\cite{HSCBHS} to replace the Ne ($1s^2 2s^2 2p^6$) core:
\be
 \sum_{i=1}^N 
    \left[
      \frac{\displaystyle\nabla_i^2}{\displaystyle 2m} + 
       V_{ext}(\bfr_i)  + 
       V^{KS}_{\lambda}(\bfr_i)
    \right]
 + \frac{1}{2} \sum_i^N\sum_{j\ne i}^N 
      \frac{\displaystyle \lambda e^2}{\displaystyle r_{ij}}.
      \label{eq:ham}
\ee
The external pseudopotential describing the interaction of the 
valence electrons with the ion core is given by
\be
 V_{ext}(\bfr_i) =  V_{loc}(r_i) + \sum_{lm} W_{l}(r_i)|lm\rangle\langle lm|, 
\ee
including a partially nonlocal term depending upon angular momentum
projectors $|lm\rangle$.
For purposes of comparison to DFT, the Hamiltonian is generalized
to consider a family of
systems characterized by the same ground-state density and
a variable coupling-constant strength $\lambda e^2$. 
A $\lambda$-dependent Kohn-Sham potential 
$V^{KS}_{\lambda}$ is added to the external potential to ensure
the invariability of the density. 
The range of interest of $\lambda$ varies from zero, 
describing the noninteracting system, for which the Hamiltonian
reduces to the Kohn-Sham equation of density functional theory,
to one, describing the physical system.  

\subsection{Variational Monte Carlo}
For XC hole expectations, we need wavefunctions 
for zero and full coupling respectively.
The $\lambda\!=\!0$ wavefunction, $\psi_0$, is 
the solution to 
Eq.~(\ref{eq:ham}) in the abscence of electron-electron coupling,
and is given by
a product of Slater determinants of single-particle orbitals.  
For the noninteracting or Kohn-Sham orbitals we take
the output of an LSD pseudopotential calculation, with the 
Kohn-Sham potential adjusted to match the spin-densities of the
$\lambda\!=\!1$ wavefunction, following the method described in 
Ref.~\onlinecite{Hood1}.

For the $\lambda\!=\!1$ wavefunction, $\psi_1$, we 
take a variational Slater-Jastrow wavefunction of the form
\be
 \psi_1 = 
    \exp\left[
            -\sum_i\sum_{j\neq i} u(\bfr_i,\bfr_j) )
       \right]
    \prod_{\sigma} D_{\sigma}\left[\phi_{i\sigma}\right],
 \label{eq:psisj1}
\ee
where $D_{\sigma}$ are Slater determinants composed of orbitals from
self-consistent LDA orbitals, before adjusting to match the VMC density.  These are close to, 
but do not exactly match the orbitals of the noninteracting wavefunction.
Interparticle correlations are described through the Jastrow prefactor
parametrized by an effective pair-potential $u$.
We use a Boys and Handy expansion of $u$~\cite{Boys,SM},
which treats 
explicitly electron-electron, electron-ion, and electron-electron-ion correlations:
\be
     u^{(N)} ({\bf r_i},{\bf r_j})
          = \!\sum_{lmn|l+m+n<N}\! C_{lmn} R_b(r_i)^l R_b(r_j)^m 
            R_b(r_{ij})^n,
     \label{eq:psisj2}
\ee
Here $R_b(r)\! =\! r / (1 + br)$ and the order of the expansion is determined
by the factor $N\!=\!l+m+n$.  Cusp conditions~\cite{Kato} are used to determine the 
coefficients $C_{lmn}$ for the linear ($N\!=\!1$) terms; otherwise terms even in 
$N$ are used and coefficients determined variationally.

This form of wavefunction is in principle exact for the Mg ($3s^2$) 
spin-singlet system, the analog of the He atom for an all-electron model,
since there are no spatial nodes in the wavefunction and the Jastrow
exponential prefactor contains all possible variables describing the 
correlation of
three charges.  For larger systems, the main source of error is likely
to be the improper determination of the nodal surface of the wavefunction.
Multiconfigurational wavefunctions in which the Jastrow prefactor is 
applied to a linear combination of Slater determinants may be useful in 
this context.  Particularly for Al and Si, nondynamic correlations in which
a ($3s^2$) spin singlet is promoted to a ($3p^2$) singlet may be important. 
At the same time, calculations for 
all-electron systems indicate that this kind of effect is important primarily
to determine the nodal structure between shells, notably, such as the 1s and 2sp
shells of Be~\cite{HarrisonHandy,FilippiJCP96,BarnettLester}, and should not be as important for a single-shell system.   

The variational Monte Carlo (VMC) method~\cite{McMillan, CepK, Foulkes}
is used to calculate expectations of the trial wavefunction and optimize
variational parameters.
The core of the method is to estimate the analytically intractable many-body
integrals that arise with the use of a Jastrow factor by integrating over a 
randomly selected set of integration points 
$R \!=\! \{\bfr_1, \bfr_2, \cdots, \bfr_N\}$ 
in the $3N$-dimensional configuration space.  These are 
sampled with a probability proportional to $|\psi_1(R)|^2$ through a 
a random walk mechanism. 
Evaluating the nonlocal pseudopotential for an integration point 
requires an additional integration over angle for each electron~\cite{FWLall,Mitas}, 
which is done here on an 18-point angular grid.
Given a set of $M$ such sample points, the energy may be estimated by the 
numerically accessible expression
\be
     \bar{E} = \frac{1}{M} \sum_i^M \psi_1^{-1}(R_i) H \psi_1(R_i).
\ee
The variational parameters are determined by the optimization of the 
variance of the energy~\cite{Umrigar} over the sample set: 
\be   
      \sigma^2 = \frac{1}{M} \sum_i^M \left[E(R_i) - \bar{E} \right]^2 
\ee
where $E(R) = \psi_T^{-1}(R) H \psi_T(R)$.
The variance is positive definite and approaches zero when the trial
wavefunction globally approaches an eigenfunction,
making for a robust
minimization process with the Levenberg-Marquardt algorithm.  

\subsection{Correlated Estimates}
%


The correlation hole is obtained from differences
between expectations of the $\lambda\!=\!0$ wavefunction $\psi_0$ and 
that of the fully coupled system $\psi_1$, differences which may be
small enough to make their detection against statistical noise difficult, 
if each expectation were calculated independently.  
Correlated estimate techniques~\cite{CepK,KalosWh},  specifically developed for
calculating arbitrarily small differences in expectations resulting from
small changes in system parameters or variational parameters, prove to be
useful here.

 
Taking the single-particle
density as an example,
the difference between the radial
densities of the two wavefunctions is obtained from the 
following expression:
\be
 \Delta n(r) = \bar{n}_1(r) - \bar{n}_0(r),
 \label{eq:corrests}
\ee
with
\be
  \bar{n}_\alpha(r)= 
  \frac{ \displaystyle\sum_k \displaystyle\frac{ |\psi_{\alpha}(R_k)|^2 }{ P(R_k) } 
\sum_i\delta(r- r_i) }
    { \displaystyle\sum_k\displaystyle\frac{|\psi_{\alpha}(R_k)|^2}{P(R_k)} },
     \;\alpha = 0, 1.
 \label{eq:correst_ni}
\ee
The $\delta$-function over the radial distance $r$ can be estimated
for a finite sampling set by a histogram method~\cite{Toulouse}.
The crucial point for the technique is that each term in 
Eq~(\ref{eq:corrests}) is summed over the same set of random 
configurations $R \!=\! \{{\bf r}_1, \ldots, {\bf r}_N\}$, 
sampled from the 
probability distribution $P(R)$.  
By using the same random walk for each case, 
the fluctuations in each evaluation become correlated
and can be partially removed in taking the difference.
In a similar fashion, the system-averaged correlation hole can be measured by
taking the correlated estimate of the difference between the pair density
expection, or intracule, of the interacting and noninteracting wavefunctions.

This  technique does not specify the form of 
probability distribution $P(R)$ to use in calculating a correlated estimate;
normally either $|\psi_1|^2$ or $|\psi_0|^2$ might be used.  Either choice
can be problematic due to undersampling -- the instance in which 
$P(R)$ goes to zero while either $\psi_1$ or $\psi_0$ remains finite.  
In this situation a rarely sampled region of space makes 
an infinite contribution $\sim |\psi|^2/P$ to an expectation.
This can occur if the two wavefunctions have 
different long-range density distributions or nodal surfaces, and can 
ultimately lead to infinite variances in the measurement of expectations.

To eliminate undersampling 
and optimize the efficiency
of the calculation, a probability distribution $P$ can be chosen 
consisting of a mixture of the probability functions for each wavefunction~\cite{CeperleyNote}.
We use
\be
 P = \left|\left(|\psi_{1}|^2 - \alpha |\psi_{0}|^2\right)\right| + 
   \epsilon\left(|\psi_{1}|^2 + \alpha |\psi_{0}|^2\right). 
 \label{mixP}
\ee
By mixing both wavefunctions together to form $P$, it becomes 
impossible for $|\psi_{i}|^2 / P$ ever to become zero -- 
should one wavefunction go to zero while the other remain finite, $P$
tends to a nonzero constant.
The choice of a difference in probabilities emphasizes areas of 
configuration space where the
difference between the two wavefunctions $\psi_1$ and $\psi_0$ is
largest, and minimizes the time spent elsewhere. 
To obtain a balanced mixture of the two states,
$\alpha$ is chosen to be roughly the ratio of the normalization
of the two wavefunctions and may be determined by measuring the expectation
of $|\psi_1/\psi_0|^2$ while calculating ground-state energies.
The small parameter $\epsilon$ is chosen to avoid reintroducing 
undersampling should $\psi_{1}$ and $\psi_{0}$ have the same value.  
With $\epsilon\!\sim\!0.1$, the method reduces the noise in 
expectations for equal sample sizes by a factor of five over non-correlated
sampling.  

%

\subsection{Calculation of exchange hole}
The approach of correlated estimates is well adapted for calculation of
changes in an expectation between two wavefunctions -- for our case,
the calculation of the correlation hole.  To calculate the expectation for
either wavefunction alone one needs an independent calculation for at least
one of the wavefunctions.  For an atom, it is straightforward to calculate the
the $\lambda\!=\!0$ expectation, the exchange hole, 
directly 
from Eq.~(\ref{eq:sysavnxsigma}).
%
%
%
For an atom with only one occupied orbital $\psi \!=\! R(r)Y_{lm}(\Omega)$ 
for a given spin, this expression gives a hole proportional to the radial
component
$|R(u)|^2$.  However for any system with two or more orbitals, 
Eq.~(\ref{eq:sysavnxsigma}) will involve convolutions between different orbitals 
and a more spread-out hole.  

This expression is the same up to choice of orbitals
as the Fermi hole, the system-averaged hole 
calculated in the Hartree-Fock approximation.
It is done here numerically, using the theoretical formalism of the 
Hartree-Fock intracule calculation of Ref.~\onlinecite{MKRD}, 
Fourier-transforming and convolving orbital pairs 
$\psi_i(\bfr) \psi^*_j(\bfr)$ 
and summing over all possible pairs.  
It proves important to keep track of 
the angular parts of the wavefunctions so that radial transforms with
high-order spherical Bessel functions are used, as well as 
angular momentum addition techniques.

\section{Results}
\label{section:results}
\subsection{Variational calculations and computational details}
Table~\ref{table:energy} shows energies and energy variances of 
variational Monte Carlo
calculations.
We show variational energies for the Silicon atom 
for the Slater determinant 
wavefunction, the Slater-Jastrow correlated
wavefunction of order $N$ [Eq.~(\ref{eq:psisj1})],
and a multideterminant-Jastrow 
wavefunction including the low energy 3s$^2$ to 3p$^2$ substitution into
the ground-state determinant
[Eq.~(\ref{eq:psisj2})].  These are compared to
configuration interaction and diffusion Monte Carlo results with 
the same form of pseudopotential as used here~\cite{Mitas}.
Considering the CI results as nearly exact, 
the bulk (92\%) of the correlation energy has already been achieved 
for the $N\!=\!4$ wavefunction. 
The quality of this wavefunction is also indicated by the 90\% reduction in 
the variance from the noninteracting wavefunction.  The most accurate wavefunction, including multiple 
determinants, misses only 2.1\% of the correlation energy, while the 
most accurate single-determinant method, the $N\!=\!8$ Slater-Jastrow,
with 24 variational parameters, misses 3.4\%.

\begin{table}
\begin{ruledtabular}
\begin{tabular}{|l|l|l|l|}
Atom &    Wavefunction &    Energy(Error) &  Variance\\
     &                 &   (hartree) &     (hartree)$^2$\\
\hline
Si &       S &               3.7188(14) &     0.188\\
Si &       SJ, N=4 &         3.8000(4) &     0.0174\\
Si &       SJ, N=8 &         3.80422(22) &    0.00943\\
Si &       CIJ, N=8 &         3.80524(28) &    0.00942\\
Si &       DMC &      3.8065(4) &\\
Si &       CI &       3.8071 &\\
\hline
Mg &       SJ, N=8 &           0.84392(4) &         0.000417\\
P &        SJ, N=8 &           6.52072(25) &         0.0187\\
Ar &       SJ, N=8 &           21.1922(7) &         0.1039\\
\end{tabular}
\end{ruledtabular}
\caption{\label{table:energy}
Optimized variational energies and variances 
for second row atoms.  The symbol SJ is for the 
Slater-Jastrow wavefunction, with the order $N$ indicated; CIJ uses
a multideterminant plus Jastrow factor, S is the Slater
determinant or $\lambda\! =\! 0$ wavefunction, DMC and CI are diffusion
Monte Carlo and configuration interaction calculations for the
same type of pseudopotential from Ref.~\cite{Mitas}.}
\end{table}

The general quality of the optimized wavefunctions across the periodic
table can be measured by the variance of the variational energy, which 
is zero for the true ground-state wavefunction.  
It is most nearly zero for the Mg atom,
which is a nodeless wavefunction and in principle exactly treated.
The variance for larger systems grows 
faster than the total energy does, but remains relatively small with the 
standard deviation in the energy for Ar about 1.5\% of the total energy.
The use of multi-configuration wavefunctions proves only to be significant 
for Si and Al.  They provide no discernible change in variational energy for 
Mg -- consistent with the SJ wavefunction being in principle exact 
for a two-electron system.
Atoms with half or more of 
the $p$-shell filled lack significant low energy configurations and 
are assumed to be described primarily by dynamic correlations.


Correlation holes are obtained using the correlated estimates technique
with $\psi_1$ taken to be the variationally optimized $N\!\!=\!\!8$ 
Slater-Jastrow wavefunction and $\psi_0$ the Slater determinant adjusted
to give the same single-particle density.
For $2\times10^5$ samples, the relative 
statistical error in the correlation hole is roughly 1\% in the 
physically relevant regime.  At long distances, as densities and thus 
Monte-Carlo samples tend to zero, the use of correlated estimates eliminate 
undersampling effectively; the worst case statistical errors are about 
100\% of the vanishingly small density differences.
For presentation in figures, data are processed with a gaussian convolution~\cite{FWLnxc}  
with a width 
of the order of three histogram bins for single-particle data 
and two bins for the pair data.  This alters the integrated correlation energy  
with a systematic upwards shift of 1 to 2\% 
but eliminates the small residual noise in the curves for better comparison
between atoms.  Energy comparisons use the unsmoothed data.

\subsection{Scaling of the valence density}
\begin{figure}
\includegraphics{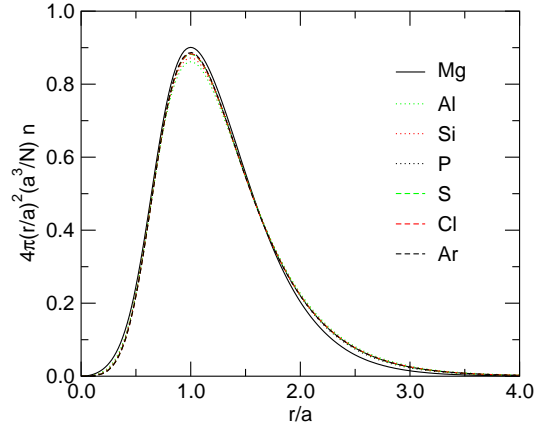}
\caption{\label{fig:density} (color online)
The scaled radial probability distribution for the
valence electrons of the second row atoms Mg through Ar,
as a function of distance in units of the radius $a$ of 
peak radial density.} 
\end{figure}

Fig.~\ref{fig:density} shows the scaled
radial probability distribution 
$4\pi r^2a n/N$ for the valence shell of 
the second row atoms from Mg ($\rm 3s^2$) to Ar ($\rm 3s^23p^6$).  
The scaling parameter $a$ is the radius at which the radial probability
distribution takes on a peak value.
The valence density across the periodic table clearly reduces to the scaling
form of Eq.~(\ref{eq:scaledens}). 
The scaling is slightly off for atoms Mg, Al, and Si with less than half-filled 
shells but is almost perfect for the other systems.
This indicates the insensitivity of the radial distribution to
the repulsive pseudopotential in the ion core region, especially for the 
larger $Z$ atoms for which the core radius shrinks rapidly relative to the
peak radius $a$.  The relative importance to the density 
of the 3s orbital, with non-zero density inside the core, 
also decreases with increasing $Z$.

\begin{figure}
\includegraphics{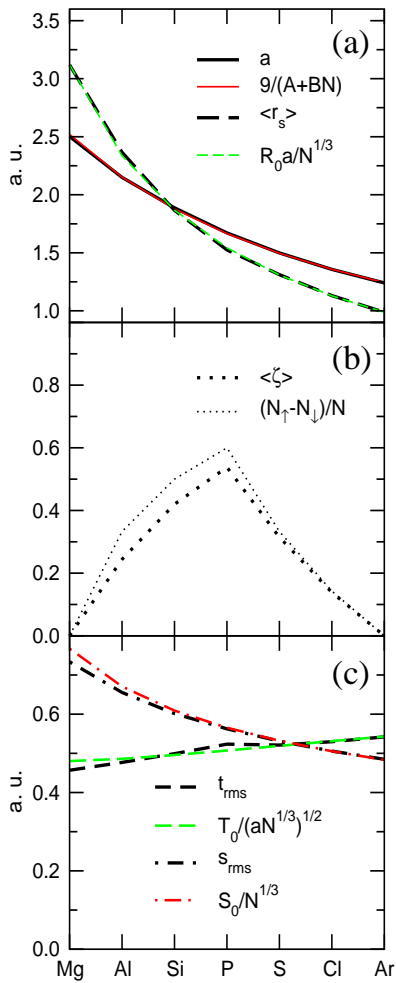}
\caption{\label{fig:avrsetc} (color online)
Density functional parameters as function of valence-electron number $N$.
Part (a) shows
the radius of peak radial density $a$ and the system-averaged 
Wigner-Seitz radius $\langle r_s\rangle$, 
(b) shows the system-averaged spin polarization $\langle\zeta\rangle$ as
compared to the fractional difference in occupation number 
$(N_\uparrow - N_\downarrow)/N$ for each spin, (c) shows the system-averaged
inhomogeneity parameters $\langle t^2\rangle$ and $\langle s^2\rangle$
as compared to scaling predictions.}
\end{figure}

The atom radius $a$ 
is plotted in Fig~\ref{fig:avrsetc}(a) versus valence number $N$; 
it gradually decreases as the number of particles
increases, indicating a rapidly increasing density.
A least squares fit to the Slater shielding model
[Eq.~(\ref{eq:slater})]
with $n^2\!=\!9$ for the 2nd row atoms yields an excellent
fit to $a$ with shielding constants 
$A\!=\!2.350(22)$ and $B\!=\!0.611(4)$.
This corresponds to a 75\% effective shielding of the nucleus from
electrons in the 3(s,p) shells by each electron in the 2(s,p) shells 
and 39\% shielding by each of the other $\rm 3(s,p)$ electrons in the 
Slater model, close to Slater's rule of thumb values of 85\% and 35\%.

Also shown in Fig~\ref{fig:avrsetc} are system averages
[Eq.~(\ref{eq:avrs})] of the parameters 
used in characterizing the semilocal density functional theory of the 
exchange-correlation 
hole.
Fig~\ref{fig:avrsetc}(a) shows the system-averaged Wigner-Seitz radius $\avrs$.
This decreases with $N$ more rapidly than the atom radius $a$, 
varying from 3 to roughly 0.9 as one goes from Mg to Ar.  
This reflects the two simultaneous effects of an increasing number of electrons 
and a shrinking atomic radius as one crosses the row.  

The average polarization $\avzeta$ shown in Fig~\ref{fig:avrsetc}(b) ranges between 
zero 
for the closed shell systems Mg and Ar, and a maximum value near 0.6 for 
P, at half-filling of the 3p shell, and varies smoothly in between.  
Although $\zeta$ is defined locally [Eq.~(\ref{eq:zeta})], its system
average approaches the natural global measure, 
$\left(N\uparrow-N\downarrow\right)/N$, also shown in Fig~\ref{fig:avrsetc}(b).
It might do this better if anisotropic densities were used in its system 
average.

The inhomogeneity factor $\avs\!=\!\sqrt{\avssq}$ for the exchange hole and
$\avt\!=\!\sqrt{\avtsq}$ for the correlation hole 
are shown
in Fig~\ref{fig:avrsetc}(c).  They show, as expected, somewhat different
scaling behavior, with $\avs$ decreasing with system size and $\avt$ fairly
constant. 
The values indicate systems with a moderate degree of
inhomogeneity. As one might expect, an isolated atom on average lies
outside of the perturbative limit $s^2 < 0.3$ characteristic
of solids, but does not quite reach the threshold of  
severe inhomogeneity, $s^2\!=\!1$, across the entire system.  

Best-fit global scaled parameters
$\tilders$, $\sqrt{\tildessq}$ and $\sqrt{\tildetsq}$ 
[Eqs.~(\ref{eq:scalers},\ref{eq:scales2},\ref{eq:scalezeta})]
are also shown in 
Fig.~\ref{fig:avrsetc}.  The fits  are weighted towards the high-$N$ end of 
the row, which shows almost perfect scaling behavior.
There is excellent agreement 
between the actual trend 
of $\avrs$ and that predicted by scaling, with $R_0\! = \!1.592(15)$.  
There is also quite good agreement between the scaling form for $\avssq$ and 
the observed value, with a falloff in quality at small $N$ because
the gradient of the density does not scale as cleanly with $N$ as does the density.
The value of $\avtsq$ shows additional behavior that follows the 
spin polarization.
This is absent if the system-average is taken over the
density of antiparallel-spin particle pairs rather than the total pair density.
Scaling values for these parameters are $S_0 \!=\! 0.969$ and $T_0\!=\!0.856$.  

\subsection{Exchange and correlation holes}
\begin{figure}
\includegraphics{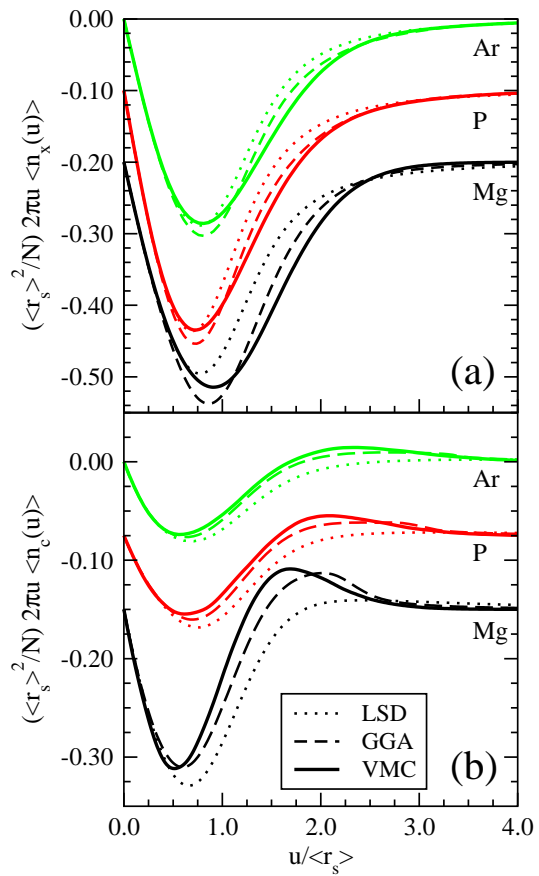}
\caption{\label{fig:xn2_combined} (color online)
Scaled and weighted system-averaged exchange hole (a), and correlation hole (b),
plotted versus scaled distance
$u/\langle r_s\rangle$ for Mg, P, and Ar.
VMC data are solid, the LSD model are dotted and the PBE GGA are 
dashed lines.  Curves for P and Mg are shifted downward for clarity; in reality they tend to zero for small and large $u$.
}
\end{figure}

Shown in Fig.~\ref{fig:xn2_combined}(a) are system-averaged 
exact exchange holes 
evaluated numerically from Eqs.~(\ref{eq:sysavnxsigma}) and~(\ref{eq:nxone}) 
and DFT models of the 
same, for three atoms: 
Mg with a filled 3s shell, P with a half-filled 3p shell, and Ar with 
a closed 3p shell.
The data for P and Mg have been shifted 
downward slightly for clarity.  
Each curve is weighted by factor of $2\pi u$ so that its integral gives the potential energy per particle
associated with the hole.  
It thus shows where the most significant contributions
to the overall energy come from.  
The holes are scaled
by a factor $\avrs^3/N$ and plotted versus the scaled interparticle distance
$\avrs$
to reflect the scaling~\cite{PW}  
of a non-polarized exchange hole under uniform scaling of the density. 
The trends for the atoms not shown are quite similar.

In comparison, the LSD hole~\cite{PW} is excellent for small $u$, out
to around the maximum of the main ``dip" that contributes the most to the exchange energy.  At the same time this main feature 
is too narrow in width.  The portion of the hole missing is spread out into
a long range tail that reflects the abrupt cutoff in occupation of states at
the fermi wavevector in a homogeneous electron gas. 
This tail contains a noticeable proportion of the total particle sum-rule
of the exchange hole (calculated by weighting the hole by $4\pi u^2$) but 
because of a negligible contribution 
(weighted by $2\pi u$) to the exchange energy, causes it to be underestimated.

The GGA hole, designed to reproduce the PBE exchange energy~\cite{EP_Xhole},
improves upon the LSD by truncating the long-range tail and 
collecting this portion of the LSD hole to form a broader hole at its 
minimum.  While failing to capture exact details, it mimics quite nicely 
on average what happens in the exact exchange hole, and thus leads upon
integration to a much
improved exchange energy.
Interestingly, the GGA is slightly less accurate than the LSD in the 
short-$u$ range of the
hole, where presumably the gradient expansion upon which the GGA is based
should be most accurate.

The LSD quite
faithfully scales with $\avrs$ -- for example, the position of the minimum in 
the curve
is the same for all three atoms.  The numerical result for the exact hole
agrees with the LSD for Ar but is shifted to larger $u$ for 
Mg.  The GGA only partly follows suit.  The reasons for this shift will
become clear in Section~\ref{section:holescaling}.


Figure~\ref{fig:xn2_combined}(b) shows the energy-weighted correlation hole for 
three cases of Mg, P, and Ar.  The hole data and interparticle 
distances are scaled and shifted in the same fashion as the exchange hole data.
The plot shows VMC data as a thick solid line, as well as the 
predictions of the LSD and the PBE GGA model~\cite{PBW_GGA}.


The correlation holes consist of 
a short-range region where the density of electron pairs  is reduced 
and a region at longer distances where it is enhanced; an overall 
sum rule of zero is required.  The length scale of the hole 
roughly follows $\avrs$, 
increasing as the number of particles decreases. 
The overall hole has a well defined finite range, with the density
removed at short range collected into a noticeable ``bump" with a maximum
at a distance between 1.33 and 2 times that of the valence shell radius $a$.  
This is intuitively reasonable
since there is little physical reason to 
enhance the pair density at interelectron distances much larger than
the diameter of the atom.  The shape of the hole varies noticeably from
more compact to more spread out as one moves across the periodic table.
Likewise, the strength of the correlation hole relative to the exchange hole 
varies considerably, with the relative strength of the Mg hole more than twice 
that of Ar.  This follows the trend in the homogeneous electron gas
from highly correlated to uncorrelated behavior as $r_s$ decreases
~\cite{OrtizBallone}.  
Unlike exchange, where
the particle sum-rule enforces more or less the same size hole in units of $r_s$, the zero sum-rule of correlation places little constraint on the 
size of the hole.

As with exchange, the LSD hole tends to predict the short
range shape of the hole quite well, with a disagreement in the 
on-top ($u\!=\!0$)
hole of 10$\%$ hidden by the $2\pi u$ weighting.  The hole
tends to be too deep and too wide.  The major disagreement
is at long distances: the HEG model for correlation
includes a long-ranged tail that screens out the long-ranged
behavior in exchange. 
To compound the effect, the system-average hole 
at long distances is dominated by contributions from the low-density asymptotic region of the atom.  These in the LSD approximation
generate unrealistically long tails to the system average.
The LSD hole for an atom ends up
predicting that the electron pair density removed at short range
is redistributed out to infinity at a rate that 
surprisingly decays even more slowly than in the HEG.  

The GGA model includes gradient corrections at short range, given
by a gradient expansion of the HEG model.  These lift up the correlation hole
at short and intermediate distances, creating a markedly better match
to the VMC results, especially for Ar.  
The 
zero sum rule for correlation is imposed 
by a finite-range cutoff of the correlation hole, which has the added
benefit of killing the long-range tails of the LSD hole.  
One therefore finds a 
consistent, systematic improvement on the LSD.
However the GGA hole dies out too slowly as compared
to the VMC at long range, so that the positive peak is too spread out and 
contributes less to the energy integral than in the VMC case. 
It is thus easier than in the case of exchange to detect the 
systematic error in the PBE model -- an overestimate of the size of the 
correlation energy.

\subsection{Exchange-correlation energy}
\begin{figure}
\includegraphics{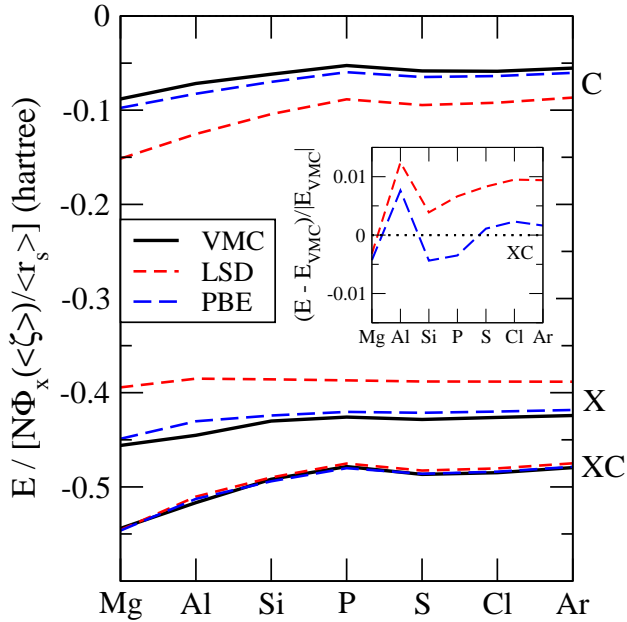}
\caption{
   \label{fig:energyanalysis} (color online)
   Exchange, correlation and exchange-correlation potential energy
   per particle for the valence shell of second row atoms, scaled by
   the factor 
   $N \phi_x(\avzeta/\avrs)$ as described in text.
   Thick solid line is VMC data; short-dashed and long-dashed lines 
   are LSD and GGA predictions.  Energy in atomic units.
   Inset: relative differences between the two DFT models and the 
   VMC data for the exchange-correlation energy.
}
\end{figure}

Fig.~\ref{fig:energyanalysis} shows correlation, exchange, and exchange-correlation energies 
for our VMC data, and for the LSD and PBE density functional models.  The 
VMC data is taken from integrating numerically the associated 
holes using Eq.~\ref{eq:exc} restricted to $\lambda\!=\!0$ 
for exchange and $\lambda\!=\!1$ for exchange-correlation.   
The density functional 
values are evaluated directly from the VMC density using 
the energy-per-particle definition conventional for DFT 
applications~\cite{JG,PBE}.
The energies are scaled by the exchange scaling 
factor of Eq.~(\ref{eq:scaleex}),
appropriate, within the LSD approximation, for a density which uniformly scales as a function of 
particle number $N$. 
This produces a nearly constant scaled value for the exchange energy
in the LSD, indicating the validity of the underlying picture.  The scaled
exchange energy $\bar{E}_x^{LSD} \!=\! -0.390(2)$~hartree may be compared to 
the homogeneous electron gas value of -0.4582~hartree, a reasonable agreement
considering the arbitrariness inherent in the definitions 
[Eq.~(\ref{eq:avrs})] of $\avrs$
and $\avzeta$.  The spin scaling of the correlation energy is not the 
same as for exchange, as demonstrated by a ``bump" at half-filling 
that correlates positively
with $\avzeta$.

Fig.~\ref{fig:energyanalysis} demonstrates the 
dramatic cancellation of errors
in the exchange and correlation components in the LSD. 
The LSD underestimates the effect of exchange, having too 
much of the sum-rule of the hole in its long-range tail, 
and overestimates the resulting
screening of this long-range tail in correlation.  
The overall error 
in the exchange-correlation energy, however, is a full order of 
magnitude smaller than that of either exchange or correlation alone. 
Having a hole derived from
an accurate many-body calculation of a true electronic system pays off in 
physically driven error cancellation.  
The PBE GGA implements a 
consistent correction of these two effects, by simultaneously creating a more
compact exchange hole and cutting off the correlation hole at long range.  
It recovers the bulk of the errors in the LSD for exchange and 
correlation separately.  
The averaged difference per electron between numerical exchange energies or
VMC-simulated correlation energies and their DFT counterparts are shown in 
Table~\ref{table:dfterror}.  The improvement from LSD to PBE is an order of magnitude for 
both exchange and correlation but more modest for the two combined.   

The major source of error in our calculation of the exact exchange energy 
is from the numerical integration of the 
exchange hole, giving errors typically less than one part in $10^5$ for the 
grid used, and is negligible here.
The VMC
results for correlation should in principle be a variational upper bound, 
assuming no numerical error in extracting them from pair density data and
the numerical calculation of exchange.  The true correlation
energy, being lower in energy, should be closer to the DFT 
predictions.  
To estimate this error in the VMC, one can compare the roughly 
three millihartree error
of our \textit{total} VMC energy for Si (Table~\ref{table:energy}), 
which we can attribute to an incomplete treatment of 
correlation, with a 20 millihartree difference between VMC and PBE correlation 
energies for Si.
On the other hand, the VMC is potentially exact for the nodeless Mg valence
 shell  -- the variational
wavefunction used converges to the exact one -- here the PBE
has a 10 millihartree error with respect to the VMC, recovering 50\% of the 
error in the LSD.  

\begin{table}
\caption{Mean absolute relative differences (mard)  and mean absolute 
difference per electron (mad), in millihartrees, between DFT models and numerical data for 
exchange (X) , correlation (C) and exchange-correlation (XC) potential energies.
\label{table:dfterror}
}
\begin{ruledtabular}
\begin{tabular}{||l|c|c|c|c|c|c||}
     & \multicolumn{2}{c|}{X} &\multicolumn{2}{c|}{C} &\multicolumn{2}{c||}{XC}\\  
\hline
     & LSD & PBE & LSD & PBE & LSD & PBE \\
\hline
\hline
mard & 0.105 & 0.017 & 0.657 & 0.117 & 0.0076 & 0.0035 \\
mad  & 28.4 & 4.7 & 26.0 & 4.6 & 2.6 & 1.0 \\
\end{tabular}
\end{ruledtabular}
\end{table}

%

The relative difference between VMC and DFT exchange-correlation energies is shown
in the inset of Fig.~\ref{fig:energyanalysis}.  It is basically on the order 
of the expected variational bias, with a mean signed relative difference 
of less than 0.01\%. 
A expected downwards shift
of 5\% in correlation energy due to variational bias will cause 
a much smaller relative change in the 
XC energy, given that correlation is a small fraction of this energy.
The shift is 0.5\% in the case of Ar; those for smaller $N$ should be similar,
involving smaller variational biases but larger relative correlation energies.
Overall, then, 
the PBE XC energy may be roughly 0.5\% higher than that of the exact 
ground-state, but systematically removing most of the LSD error.


%

One may take the exchange scaling analysis  one step further to analyze the 
gradient contributions
to the exchange and correlation energies.  In particular, the GGA predicts
a multiplicative correction to the LSD exchange energy-per-particle of
\be
    \epsilon_x^{GGA} / \epsilon_x^{LSD} \sim 1 + \mu s^2,
    \label{eq:fxgga}
\ee
for small values of $s^2$.
The PBE data in Fig.~\ref{fig:energyanalysis}
shows a slight variation from LSD scaling for smaller $N$ or larger $\tildessq$
which can be fit to Eq.~(\ref{eq:fxgga}). 
The value of $\mu$ thus obtained is $0.351$ for 
the GGA data as compared to $0.431$ for the exact calculated exchange, 
a 20\% stronger response to inhomogeneity than the GGA prediction.
\subsection{Scaling trends of exchange and correlation holes}
\label{section:holescaling}
\begin{figure}
\includegraphics{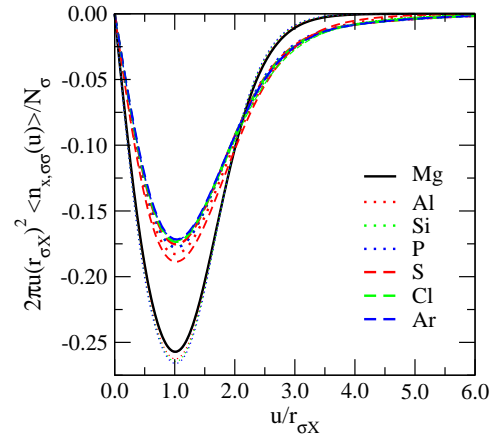}
\caption{
   \label{fig:xn2xscaled} (color online)
   Scaled minority-spin (thinner lines) and majority-spin (thicker lines) 
   exchange holes for second row 
   pseudopotential atoms plotted versus a spin-dependent, scaled interparticle 
    distance 
   $u/r_X^\sigma$.  
   Each hole is scaled by $(r_X^\sigma)^3$ and weighted by $2\pi$ times the scaled 
   interparticle distance.
   The scaling length $r_X^\sigma$ is determined by the distance at which the 
   weighted hole has its minimum value.
}
\end{figure}

Under a uniform scaling of the density -- and constant particle number $N$ --
the exchange hole is invariant.
As our densities scale
fairly uniformly, but with $N\!=\!Z$ not constant, it is informative to see 
what extent the exchange hole can be reduced to a scaling form.
This is most easily done by considering
the spin-decomposed exchange hole, Eq.~(\ref{eq:sysavnxsigma}). 
To do so, we use an identifiable point of each hole 
-- the minimum of the 
energy-weighted hole -- to determine a length scale $r_X^\sigma$ for 
each spin species $\sigma$.  The hole 
is normalized by the 
number of particles of that spin to guarantee a sum rule of -1.
Uniform scaling then
entails the same procedure as for the radial probability density: $n_x^\sigma$ 
scales to $(r_X^\sigma)^3 n_x^\sigma$ as distance is scaled from $u$ to $u/r_X^\sigma$.
The results are shown in Fig.~\ref{fig:xn2xscaled}.

We do see scaling behavior, but interestingly, \textit{two} scaling forms.
There is a striking difference between the
hole formed from a single particle, (in Mg, and the minority spin channel of
Al through P) and that of two or more particles.
The 
two cases that disagree slightly from this trend are, quite naturally, the
two cases with only two electrons in a given spin channel, Al and Si.
Otherwise the results neatly scale on top of each other.  

This result is not that surprising if we consider the 
form of the exchange hole in Eq.~(\ref{eq:sysavnxsigma}).  For one particle
in a given spin, the form reduces to a convolution of the spin density
$n_{\sigma}(\bfr) \!=\! |\psi_{3s}^{\sigma}(\bfr)|^2$ and leads to a relatively 
compact function.  
The hole merely removes 
the self-interaction error of the Hartree approximation, and in a sense
is not a true \textit{exchange} hole.  
The $N_\sigma\! >\!1$ case also includes 
convolutions of the overlap of 
two different orbitals 
$\psi^*_{i,\sigma}(\bfr)\psi_{j,\sigma}(\bfr)$, that describe the exchange
of two electrons.
Such overlap terms are naturally more spread out than a single-orbital 
probability 
 and create a more slowly decaying hole. 
Note that the exchange scaling law [Eq.~(\ref{eq:exscaling})] requires that both forms scale 
uniformly with an isolectronic (fixed $N$)
uniform scaling of the density, but with significantly different asymptotic
forms because of their different origins.
Finally it is interesting to note how quickly the transition from
the self-interaction dominated to an exchange dominated hole occurs -- 
the large number limit is essentially reached starting at $N_{\sigma}\!=\!2$.

\begin{figure}
\includegraphics{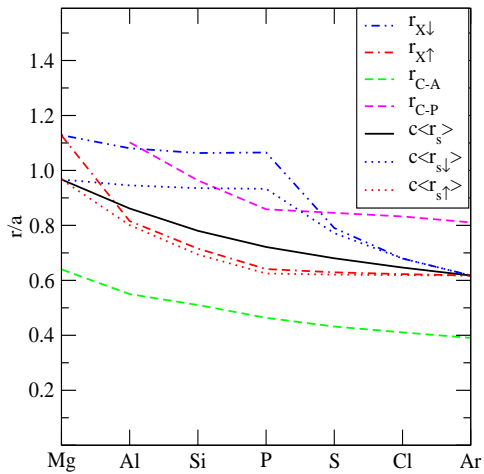}
\caption{
\label{fig:rscales} (color online)
Various scaling lengths discussed in the paper.  Shown versus valence
electron number are: the position of maximum depth ($r_X$) in the 
spin-up (red) and spin-down (blue) energy-weighted exchange holes, 
the same ($r_C$) for antiparallel (green) and parallel spin correlation holes, 
all scaled by the effective valence bohr radius $a$. 
System-averaged Wigner-Seitz radius $\avrs$ for the total density 
(black) and each spin density (dotted lines) are also shown, scaled by a factor
of $c\!=\!0.755$ to aid in comparison.
}
\end{figure}

Fig.~\ref{fig:rscales} shows scaling lengths 
for the majority (up) spin and minority (down) spin exchange holes, 
$r_X^{\uparrow}$ and $r_X^{\downarrow}$,
in units of the valence-shell radius $a$ for each atom in the
second row.  
In addition, the average Wigner-Seitz radius $\avrs$ is plotted 
and the equivalent spin-dependent
radii $\avrssigma = \avrs (1 + \sigma\zeta)^{-1/3}$, proportional to the
natural length-scale $(k_F^\sigma)^{-1}$ of the spin-decomposed 
LSD exchange hole.
These are scaled by a factor of 0.755 so that they match $r^\sigma_X$ for
Ar.  The comparison between the actual length-scales $r_X$ and the LSD 
equivalent shows the separation of the single-particle
and multi-particle cases seen in Fig.~\ref{fig:xn2xscaled}.
For spin occupation $N_\sigma >1$, 
the $r_X$ values are well predicted by LSD
theory.  For $N_\sigma\!=\!1$ 
the hole scales as $a$, and is notably
larger than the LSD prediction.  In this case, the length scale is set 
simply by the width $\sim a$ of the single orbital occupied.

Unlike exchange, the universal scaling law for 
correlation~[Eq.~(\ref{eq:ecscale})] is nontrivial,
and the correlation energy and hole are nontrivial to model even for 
the homogeneous electron gas.  
Nevertheless, Fig.~\ref{fig:xn2_combined}(b) shows that correlation holes for 
the second row atoms are qualitatively similar and it is instructive to 
scale the holes to highlight the trends that occur as one crosses the row.

\begin{figure}
\includegraphics{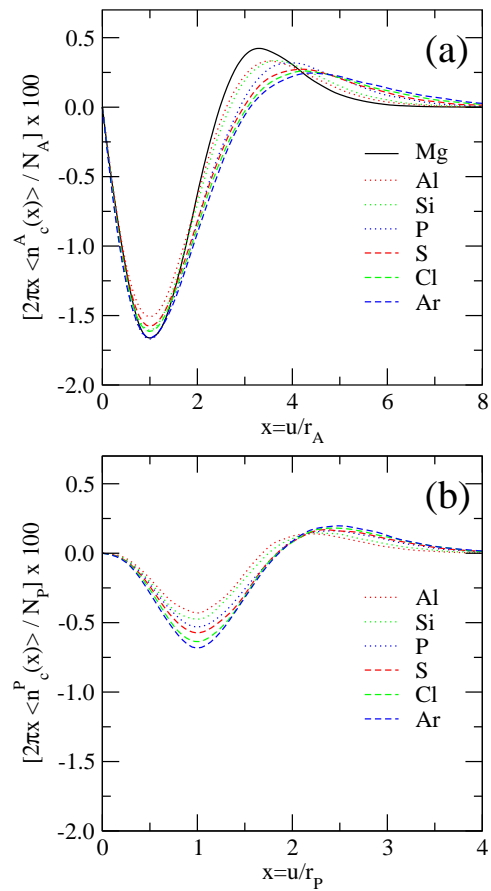}
\caption{
   \label{fig:xn2cscaled} (color online)
   Scaled correlation holes for second row pseudopotential atoms.  
   Part (a) shows the correlation hole at full coupling for antiparallel-spin 
   electrons, weighted by $2\pi$ times the interelectron 
   distance $u$.  The holes are scaled by the number of antiparallel electron 
   pairs $N_A$ and plotted versus the scaled electron distance $x=u/r^A$,
   where $r^A$ is the distance at which each weighted hole has
   a minimum.  Part (b) shows the analog for the parallel-spin 
   correlation hole, in terms of parallel-spin pair number,
   $N_P$ and scaling radius $r^P_C$.  Each hole is plotted in atomic units. 
}
\end{figure}

To do this we use an empirical scaling relation that matches the holes 
as closely as we find possible.  
It is again helpful to spin-decompose
the holes, into antiparallel-spin (A) and parallel-spin (P) channels.  
A scaled form $\ncbar(x)$ of the hole in either channel may be constructed
as
\be
     2\pi u \langle n_{c}^{A,P}(u) \rangle =
            2\pi (u/r^{A,P}_C) N_{A,P}\; \ncbar^{A,P}(u/r^{A,P}_C),
     \label{eq:ncscale}
\ee
where $N_{A}$ and $N_{P}$ are the number of electron pairs of either 
spin channel and $r^A_C$ and $r^P_C$ are 
scaling lengths.
These are again chosen   
as the distance at which the energy-weighted holes take their
minimum value.  This optimizes the match between holes at short
interparticle distances at the expense of that at longer ones.
The results 
are shown as a function of scaled interparticle distance in 
Fig.~\ref{fig:xn2cscaled}; scaling lengths are shown in Fig.~\ref{fig:rscales}.

For the case of antiparallel-spin electron pairs [Fig.~\ref{fig:xn2cscaled}(a)], the results 
at short distances match up closely.  The minimum of the holes 
shows some shell structure, with the closed spin-shell atoms 
Mg, P and Ar with the deepest holes.  At
long range, there is a systematic trend 
as the $3p$-shell is filled, 
going from a compact 
hole with a sharp positive peak, to a relatively
wider hole with a positive peak spread out over a large range
of distance relative to $r^A_C$.  The trend seems to be one of gradual 
reduction of finite-size effects at long-range, with the shape of the 
hole trending to that of the homogeneous electron gas, with 
its infinite-ranged hole.

The parallel-spin holes are shown in Fig.~\ref{fig:xn2cscaled}(b).  That for
Mg is trivially zero since there are no same-spin pairs. 
As shown in Fig.~\ref{fig:rscales}, $r^P_C$ is almost exactly twice that of $r^A_C$ across the second row so that each subplot of Fig.~\ref{fig:xn2cscaled} 
shows the same physical range in distance for each atom despite the different 
abscissas.
The hole-per-pair is significantly smaller than in the antiparallel spin channel
and is concentrated at longer interparticle distance.
Both effects are caused by the exchange hole which removes 
most of the probability of finding particles of the same spin at short range.
As one goes across the shell, the tendency
is for gradually deeper and longer-ranged holes.  
Finally, given the equivalence $r^P_C\!=\!2r^A_C$, one finds that the scaled 
antiparallel and parallel holes-per-pair for each
atom match fairly closely at large distances ($r^A_C\! >\! 5$).  
At distances longer than the effective exchange hole radius, 
it seems that electrons no longer detect each other's spins.

All-electron spherically-averaged Coulomb hole 
data
for nearly spherical molecules and atoms show
qualitatively very similar results to ours, after taking into account the 
$4\pi r^2$ weighting typically used in the 
literature~\cite{BanyardMobbs, WTS2, Galvez:034302,*Galvez:044319, Toulouse}.
The hole at small interparticle distances typically shows the evidence of 
shell and molecular structure, but significantly reduced 
by the effects of system averaging, particularly in the large-$Z$ limit.   
The qualitative shape of the peak at large 
interparticle distances seems to be a universal feature of 
the Coulomb/correlation hole
and closely matches the behavior seen in Figs.~\ref{fig:xn2_combined}(b) and~\ref{fig:xn2cscaled}.
Shell analysis of the hole for small atoms~\cite{BanyardMobbs} tells us 
that this feature is caused by electrons in the valence shell
and is thus apropos for comparison to our data.  
Using the point of crossover $r_{cr}$ from negative hole to positive peak 
as a point of reference, we find that our scaling model provides a 
fairly good prediction for other data.  
A fit of this point to our data yields $r_{cr}=1.5*R_0a/N^{1/3}$ for Si; 
using this formula and the naive Slater model for $a$ 
we predict Coulomb hole crossover points for 
the C isolectronic series that range from 
11.2 a.u. for C to 8.3 a.u. for Ne$^{+4}$, which are indistinguishable from 
those of the Coulomb hole plots of VMC data for the series~\cite{Galvez:034302,*Galvez:044319}.
This suggests that our data may be useful to analyze the long-range
trends of correlation holes of atoms and small molecules. 

A few extra notes are in order -- first of all, the 
point of comparison between holes found to be most successful 
is the correlation hole 
\textit{per pair} and not per particle, as with exchange.  
A particle number for the antiparallel-spin hole cannot be unambiguously
defined, and we find no scheme for a per-particle hole that provides as 
consistent a scaling fit for the series.  At the same time, the scaled hole
used here is 
not scale-invariant, in which case the hole, having
units of volume, should scale as $1 / r_{scale}^3$.  
It is therefore not unitless, but rather varies as $1/a_0^2$ given the
choice of atomic units. 
The situation is reminiscent of that of the slowly
varying electron gas, in which the inhomogeneity in the exchange hole can be
described by the scale-invariant parameter $s$, but the correlation hole 
depends on the non-scale-invariant $t \sim s/\sqrt{r_s}$.
However, it is not impossible
to scale the correlation hole with a uniform scaling hypothesis of the 
form:
\be
      \sysnc = \frac{F(N_\uparrow,N_\downarrow)}{(r_{scale})^3} 
               \bar{n}_c(u/r_{scale})
\ee
for an arbitrary function $F(N_\uparrow,N_\downarrow)$ of the number 
of valence electrons of each spin.  
In this case, the absence of scaling behavior is assumed to lie in the 
arbitrary amplitude $F$.  The two approaches may be considered equivalent
since the normalization used in Eq.~(\ref{eq:ncscale}) is implicitly a function 
of $N_\uparrow$ and $N_\downarrow$.

It is also worth noting that length-scales for
correlation and exchange diminish as a proportion of 
atom radius $a$ in going from Mg to Ar~(Fig.~\ref{fig:rscales}).  
Finite size effects in exchange and correlation
become important as $\avrs/a$
approaches unity -- that is, the length-scale of the XC hole approaches
the system size of the atom.
%
We thus expect, and
find, the largest errors for local DFT's for Mg, for which $\avrs/a$ is
largest, and the smallest for Ar.
Nevertheless, the 
GGA parameter $t^2$ used to estimate this inhomogeneity error has 
a system average that \textit{increases} as one proceeds down the row.
In the slowly varying electron gas, the higher the electron
density, the more sensitive correlation is to inhomogeneity, but in atoms,
the higher the density, the less effect the finite size of the atom has
on correlation.
This misidentification is a natural limitation of using a semilocal parameter 
to measure the effect of inhomogeneity, which for atoms as essentially
zero-dimensional objects must necessarily 
depend on global features.   Interestingly, however, the PBE correlation 
hole does take into account one global measure, the zero particle-sum-rule 
of the hole.  For very inhomogeneous systems, this constrains the hole
to react more strongly at larger $r_s$ to a given value of $t^2$.
Thus it follows to a fair degree the observed finite-size effects, 
more so than one might have expected.

\subsection{Pair correlation function}
One can gain more insight into the behavior of correlation with atomic
number, and in particular the role of finite size effects, by calculating a
pair correlation function
$g_c$, defined as the ratio of the pair density of the fully
correlated ($\lambda \!=\! 1$) system to that of the equivalent 
noninteracting system ($\lambda \!=\! 0$):  
\be
g_c(u) = \frac{\langle n^{(2)}_{\lambda=1}(u)\rangle}
{\langle n^{(2)}_{\lambda=0}(u)\rangle}.
\label{eq:gc}
\ee
This maps the fractional
change, 
because of Coulomb correlation, 
in the expected number of particle pairs as a function of distance. 
As the noninteracting-system pair density already incorporates exchange, 
this measures a purely correlation effect, 
and is somewhat different from the normal definition of $g$
measured relative to the pair density of independent
particles.   
The value at zero separation, $g_c(0)$, measures the
on-top hole for antiparallel spin correlation. 
Because of fermi statistics, the parallel-spin channel 
contributes zero to both numerator and denominator of the on-top 
value of Eq.~(\ref{eq:gc}).  

\begin{figure}
\includegraphics{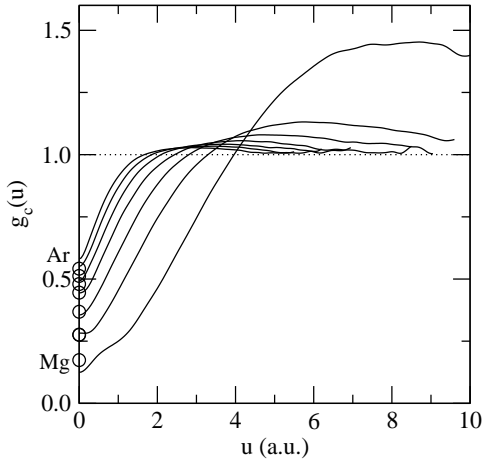}
\caption{
   \label{fig:gch2av}
    The correlation contribution to the pair correlation function, $g_c$, 
    for the valence shell of the second row atoms plotted versus interparticle distance $u$.
    The curve for Ar and Mg are labelled at $u\!=\!0$.  Also shown 
(circles) is the system-averaged 
    on-top or $u\!=\!0$ pair-correlation function for the LSD. 
}
\end{figure}

We show the VMC values for 
$g_c(u)$ for the second row atoms in Fig.~\ref{fig:gch2av}.
As a comparison, the equivalent system-averaged on-top pair correlation
function for the LSD, 
$\langle g^{LSD}_c(0)\rangle$, calculated by the same 
system-averaging technique as Eq.~(\ref{eq:avrs}), is shown as circles
at $u\!=\!0$.
At short range the pattern of the VMC data is similar to the 
correlation hole in the 
HEG~\cite{OrtizBallone}
except for deviations, due to small-number statistics, 
from an expected cusp~\cite{Kimball} 
at very short interparticle distances. 
Given the noise in the VMC in this region, 
the LSD and VMC values for the on-top hole 
are in reasonable agreement.
At long range one sees a marked enhancement of the
distribution of particle pairs relative to the uncorrelated pair density.
The long-range asymptotic value of the pair distribution function in the HEG 
is one, indicating a vanishing difference between correlated 
and uncorrelated distributions. 
In atoms, this is not a 
requirement, and in fact the enhancement of pairs at long range
is surprisingly large for Mg.  
The distances in this case are bigger than the distance
from the peak of the valence density on one side of the atom to that on the
other, about 1.7~$a$ or roughly 3.5~a.u.\ for Mg.
There is a large fractional enhancement of the relatively small density of 
pairs separated by several atomic radii, in contrast to 
the modest enhancement of the long-range pair density for more
localized holes.
This is consistent with an ``in-out" correlation where 
if one electron is found on the inside of the valence shell the other 
is favored to be found on the outside edge of the shell.

\section{Discussion}
\label{section:conclusions}
The valence shell of first or second row atoms, within the pseudopotential 
approach, is an example of uniform scaling that has been known from the earliest
stages of atomic physics.  
The form of scaling is 
related to that of the Thomas-Fermi scaling of the all-electron density
for large atoms, in that the net charge of the system is kept fixed as 
the system is scaled, and so scaling parameters such as  $\avrs$ depend upon 
the number of electrons $N$.  It differs in that the scaling parameter $a$ 
does not have a simple power-law behavior with $N$, but essentially a 
Pade-like dependence, as noted by Slater, because of the role of 
self-interaction.  
At the same time, the 
scaling of a single shell is a form of uniform scaling 
to high density, and might be a useful complement to the standard example
of isoelectronic scaling.  What happens in the current case, in the limit of a 
very large-degeneracy shell, for which $N\gg 1$ is achievable?  

Intriguingly, the exchange hole has not one, but two, scaling forms -- with 
the single-orbital hole fundamentally different from those that are 
constructed with two or more orbitals.  Self-interaction effects spoil
the invariance of the exchange hole under uniform scaling with constant
net charge that would hold with constant particle number.
But at the same time, self-interaction effects die out astonishingly rapidly,
with systems of three or more electrons already showing large-$N$\
scaling behavior.  
Correlation holes do not scale uniformly and trends across the second row 
cannot be unambiguously modeled.   Nevertheless, a clear 
trend with the scaling of the density does occur: the transition from more correlated,
low density systems in which finite size effects dominate
the correlation peak at large interparticle distances, 
to higher density ones in which the system-averaged hole approaches that
of the HEG, and the GGA approximation is very good. 
The natural parameter to characterize this trend, the ratio $\avrs/a$ of 
average hole radius to atomic radius, can not be modeled in a semilocal DFT,
but the use of a different constraint -- a cutoff based on the correlation hole
sum rule -- does a reasonable job of following it.
The qualitative behavior of the positive correlation peak is seen in 
all-electron calculations and a crude estimate indicates that scaling
results for this feature should apply to atoms and spherical molecules
in general.

It is of interest to analyze the sources of error in the PBE correlation hole 
by decomposing the system averaging.  The PBE hole near the 
valence peak determines the overall shape of the system-averaged hole but,
since gradients are small in this region,  
 has an unphysical long-range tail equivalent to that of the LSD hole.
Holes in the pseudopotential core and far from the atom
deviate dramatically from the system average but cut 
off just at the right length scale.  
They end up contributing the most to the positive peak 
of the system-average that, physically, is caused by finite size of the atom.  
It is not the local XC hole but its system-average that the PBE is capturing,
as it was designed to do.

Our data thus show that for the type of system studied here, the semilocal 
PBE GGA model works essentially as advertised.  The significant defects of 
the XC hole in the LSD approach are more or less fixed, especially for the 
crucial aspect of the hole -- the integral that produces the exchange and 
correlation energy.  The reason for this success is likely related to the 
simplicity of the single-shell structure studied, with the main physics being 
scaling behavior similar to that which the PBE was built to represent.  
The main source of 
error, the poor treatment of finite-size effects, occurs mainly in the 
long-range tail of the GGA XC hole which does not contribute much to the 
total energy.

The PBE does somewhat underestimate the gradient 
correction parameter
$\mu$ [Eq.~(\ref{eq:fxgga})] needed for valence-shell exchange energies.
Recent work on modifying the PBE for solids~\cite{MattssonGGA,PBEsol,SOGGA} indicates that use 
of a value for $\mu$ half that of the PBE, but consistent with the gradient
expansion for a slowly-varying electron gas,
leads 
to improved lattice constants and bulk moduli (if poorer cohesive 
energies).  The PBE choice, is instead best suited for predicting total 
energies of atoms and binding energies of molecules.  
Our work thus emphasizes the incompatibility between 
GGA's designed for molecules and those for solids.
The large value of $\mu$ needed for total atomic energies
has been attributed~\cite{PCSB-PRL97} to using the gradient correction to
account for exchange energy corrections caused by
the cusp in electron density near the nucleus.
This is should not be true in the present case since the nucleus has been
replaced by a smooth pseudopotential.
It seems that another mechanism is at play here, quite possibly 
self interaction.  The PBE models 
systems with large self interaction error, like the two-electron valence shell 
of Mg, very well.  But this is perhaps at a cost of overcorrecting in
situations like the slowly varying electron gas where inhomogeneity and
not self-interaction is the predominant issue.  Another consideration is
the boundary condition difference between finite and infinite systems, with
the former needing a stronger correction to produce a more sharply defined 
long-range cutoff to the hole.

Our work also points to the difficulty of imposing  a self-interaction
correction to the GGA.
To the extent that the GGA may be correcting LSD error caused by 
self-interaction
and not by the imperfect treatment of inhomogeneity, applying a 
self-interaction correction     
to the GGA would lead to correcting the same problem twice.
Any self-interaction correction based on a GGA would 
require a remarkable degree of cancellation of errors between the 
correction for exchange and that for correlation to improve total energies
for the systems studied here.  
The exploration of self-interaction corrected GGA models that could have this
level of error cancellation might thus be of interest.

\section{Acknowledgments} 
One of us (ACC) thanks John Perdew and Cyrus Umrigar for 
helpful discussions and comments.


\begin{thebibliography}{79}
\bibitem [{Adi()}]{AdiabaticFootnote}%
  \BibitemOpen
  \href@noop {} {}\bibinfo {note} {It is of interest to note that the variation
  of the hole for a given physical system with coupling constant $\lambda$ is
  equivalent to the uniform scaling of the density at fixed coupling constant.
  This would be a useful complement to the study we are carrying
  out.}\BibitemShut {Stop}%
\bibitem [{Sla()}]{SlaterVirial}%
  \BibitemOpen
  \href@noop {} {}\bibinfo {note} {The presence of self-interaction error in
  the Hartree energy is what forces the constant term $A$ in this expression,
  and spoils a purely power-law behavior as seen in Thomas-Fermi scaling of the
  all-electron density.}\BibitemShut {Stop}%
\bibitem [{foo()}]{footnoteUc}%
  \BibitemOpen
  \href@noop {} {}\bibinfo {note} {The analogous relationship for the
  correlation hole is used to derive the scaling formula for the hole at full
  coupling [Eq.~(\ref{eq:gcscaling})].}\BibitemShut {Stop}%
\end{thebibliography}

%

%
\end{document}